\documentclass[preprint,12pt]{aastex}

\usepackage{graphicx}
\usepackage{wasysym}
\usepackage{natbib}

\begin{document}

\newcommand{\hide}[1]{} 
\thispagestyle{empty}

\begin{center}
\LARGE
NOTE: Ridge Formation and De-Spinning of Iapetus via an Impact-Generated Satellite
\vspace{2 in}

\normalsize
H. F. Levison$^1$, K. J. Walsh$^1$, A. C. Barr$^1$, L. Dones$^1$

\vspace{0.5 in}
$^1$Southwest Research Institute, 1050 Walnut St. Suite 300, Boulder, CO 80302, USA

\vspace{0.5 in}
{\tt Corresponding Author:}\\ 
Harold F. Levison\\ hal@boulder.swri.edu \\
Southwest Research Institute\\
1050 Walnut St. Suite 300\\ 
Boulder, CO 80302, USA\\
1 (303) 546-0290 \\

\end{center}

\newpage

\section*{Abstract}

We present a scenario for building the equatorial ridge and
de-spinning Iapetus through an impact-generated disk and satellite.
This impact puts debris into orbit, forming a ring inside the Roche
limit and a satellite outside.  This satellite rapidly pushes the ring
material down to the surface of Iapetus, and then itself tidally
evolves outward, thereby helping to de-spin Iapetus.  This scenario
can de-spin Iapetus an order of magnitude faster than when tides due
to Saturn act alone, almost independently of its interior geophysical
evolution. Eventually, the satellite is stripped from its orbit by
Saturn.  The range of satellite and impactor masses required is
compatible with the estimated impact history of Iapetus.

\section{Introduction}

The surface and shape of Iapetus (with equatorial radius,
$R_{\mathrm{I}}$=746 km, and bulk density, $\bar{\rho} =
1.09$~g~cm$^{-3}$) are unlike those of any other icy moon
\citep{Jacobson2006}.  About half of Iapetus' ancient surface is dark,
and the other half is bright \citep[see][for discussion]{Porco2005}.
This asymmetry has been explained recently as the migration of water
ice due to the deposition of darker material on the leading side of
the body \citep{Spencer2010}.  Iapetus also has a ridge system near
its equator, extending $>110^{\circ}$ in longitude \citep{Porco2005},
that rises to heights of $\sim$ 13 km in some locations
\citep{Giese2008}.  The ridge itself is heavily cratered, suggesting
it originated during Iapetus' early history.  Finally, Iapetus'
present-day overall shape is consistent with a rapid 16-hour spin
period rather than its present 79-day spin period
\citep{Thomas2010,CR2007, Thomas2010}.

To some, the equatorial position of the ridge and Iapetus' odd shape
suggest a causal relationship.  Most current explanations invoke
endogenic processes.  For example, detailed models of Iapetus' early
thermal evolution suggest that an early epoch of heating due to
short-lived $^{26}$Al and $^{60}$Fe is required to close off
primordial porosity in the object while simultaneously allowing it to
rapidly de-spin, cool, and lock in a ``fossil bulge'' indicative of an
early faster spin period \citep{CR2007, Rob10}.  Recently, Sandwell \&
Schubert (2010) suggested a new and innovative mechanism for forming
the bulge and ridge of Iapetus through contraction of primordial
porosity and a thinned equatorial lithosphere.  However, only a narrow
range of parameters allows both a thick enough lithosphere on Iapetus
to support the fossil bulge, while also being sufficiently dissipative
to allow Iapetus to de-spin due to Saturn's influence on solar system
timescales.

In these scenarios, the ridge represents a large thrust fault arising
from de-spinning.  One difficulty faced by these ideas is that the
stresses arising from de-spinning at the equator are perpendicular to
the orientation required to create an equatorial ridge
\citep{Melosh1977}. Other interior processes, such as a convective
upwelling \citep{CLK2008}, or convection coupled with tidal
dissipation driven by the de-spinning \citep{RobertsLPSC2009} are
required to focus and reorient de-spinning stresses on the equator.
These latter models have difficulty reproducing the ridge topography
because thermal buoyancy stresses are insufficient to push the ridge
to its observed height \citep[see][]{DombardLPSC2008}.

Alternatively, the ridge may be exogenic. One leading hypothesis is
that the ridge represents a ring system deposited onto Iapetus'
surface \citep{Ip2006,DombardAGU2010}.  This model has the benefit of
providing a natural explanation for the mass, orientation, and
continuity of the ridge, which present a challenge to endogenic
models.

Here we extend this idea to include a satellite that accretes out of
the ring system beyond the Roche limit.  As we show below, this can
significantly aid in the de-spinning of Iapetus.  In particular, we
hypothesize that:

\noindent 1) Iapetus suffered a large impact that produced a debris
disk similar to what is believed to have formed Earth's Moon
\citep{Canup2004,Ida97,Kokubo00}. Like the proto-lunar disk, this disk
straddled the Roche radius of Iapetus, and was quickly collisionally
damped into a disk.  As a result, a satellite accreted beyond the
Roche radius, while a particulate disk remained on the inside.  Also,
the impact left Iapetus spinning with a period $\leq 16\,$hr, thereby
causing the bulge to form.\footnote{It is important to note that the
  impact that we envision is in a region of parameter space that has
  yet to be studied. Such an investigation requires sophisticated
  hydrodynamic simulations and thus is beyond the scope of this paper.
  We leave it for future work. We emphasize, however, that the general
  geometry we envision has been seen in many hydrodynamic simulations
  of giant impacts \citep[e.g.,][]{Canup2004}, so we believe that our
  assumed initial configuration is reasonable.}

\noindent 2) Gravitational interactions between the disk and Iapetus'
satellite (hereafter known as the sub-satellite) pushed the disk onto
Iapetus' surface, forming the ridge.  As Ip (2006) first suggested, a
collisionally damped disk, similar to Saturn's rings, will produce a
linear feature precisely located along the equator.  Thus, it
naturally explains the most puzzling properties of the ridge system.
The impact velocity of the disk particles would have been only
$\sim\!300\,$m s$^{-1}$ and mainly tangential to the surface, so it is
reasonable to assume that they would not have formed craters, but
instead piled up on the surface.

\noindent 3) Tidal interactions between Iapetus and the
sub-satellite led to the de-spinning of Iapetus as the sub-satellite's
orbit expanded.  Eventually, the sub-satellite evolved far enough from
Iapetus that Saturn stripped it away. Iapetus was partially de-spun
and continued de-spinning under the influence of Saturn.  Finally, the
sub-satellite was either accreted by one of Saturn's regular
satellites (most likely Iapetus itself) or was ejected to heliocentric
orbit (cf$.$ \S{\ref{sec:impact}}).  The end-state is a de-spun
Iapetus that has both a bulge and a ridge.  Faster de-spinning aided
by the presence of a sub-satellite likely relaxes constraints on the
early thermal evolution of Iapetus determined by prior works
(Castillo-Rogez et al~2010; Robuchon et al.~2010).

Because the results of one part of our story can be required by other
parts, we begin our discussion in the middle and first find, through
numerical simulations, the critical distance ($a_{\mathrm{st}}$) at
which a sub-satellite of Iapetus will get stripped by Saturn. Knowing
this distance, we integrate the equations governing the tidal
interactions between both Saturn and Iapetus, and between Iapetus and
the sub-satellite, to estimate limits on the mass of the
sub-satellite.  We then study the fate of the sub-satellite once it
was stripped away from Iapetus by Saturn.  Finally, using crater
scaling relations we reconcile a sub-satellite impact with the
topography of Iapetus.

\section{Satellites Stripped by Saturn}

The distance at which a satellite of Iapetus becomes unstable is
important for calculating tidal evolution timescales.  In systems
containing the Sun, a planet, and a satellite, prograde satellites are
not expected to be stable beyond $\sim$~$R_{\mathrm{H}}$/2, where the
Hill radius is defined as $R_{\mathrm{H}}=a(m/3M)^{1/3}$ with $a$ as
the planet's semi-major axis, $m$ as its mass, and $M$ as the total
system mass \citep{Hamilton91, Barns02, Nesvorny2003}.  In our case,
Iapetus plays the role of the planet, and Saturn the role of the Sun.
However, the tidal evolution timescale depends strongly on semi-major
axis (as the $-13/2$ power, Eq$.$~\ref{dadt}) and thus the success of
our model depends sensitively on the value of the critical distance,
$a_{\mathrm{st}}$.  Therefore, we performed a series of numerical
simulations to determine $a_{\mathrm{st}}$.

This experiment used the swift\_WHM integrator (Levison \& Duncan
1994; which is based on Wisdom \& Holman
1991\nocite{Levison94,Wisdom1991}) to integrate two sets of test
particles consisting of 500 objects, each of which were initially on
orbits about Iapetus with semi-major axes, $a$, that ranged from
0.1--0.8~$R_{\mathrm{H}}$. The particles in the first set were
initially on circular orbits in the plane of Iapetus's equator.
Particles in the second set had initial eccentricities, $e$, of 0.1,
and inclinations, $i$, that were uniformly distributed in $\cos{(i)}$
between $i\!=\!0$ and $i\!=\!15^\circ$. Saturn is by far the strongest
perturber to the Iapetus-centered Kepler orbits and is the main source
of the stripping.  For completeness, we have also included the Sun and
Titan.  The effects of the other Saturnian satellites are at least two
orders of magnitude smaller than those of Titan and thus can be
ignored.

\begin{figure}
\includegraphics*[width=6in]{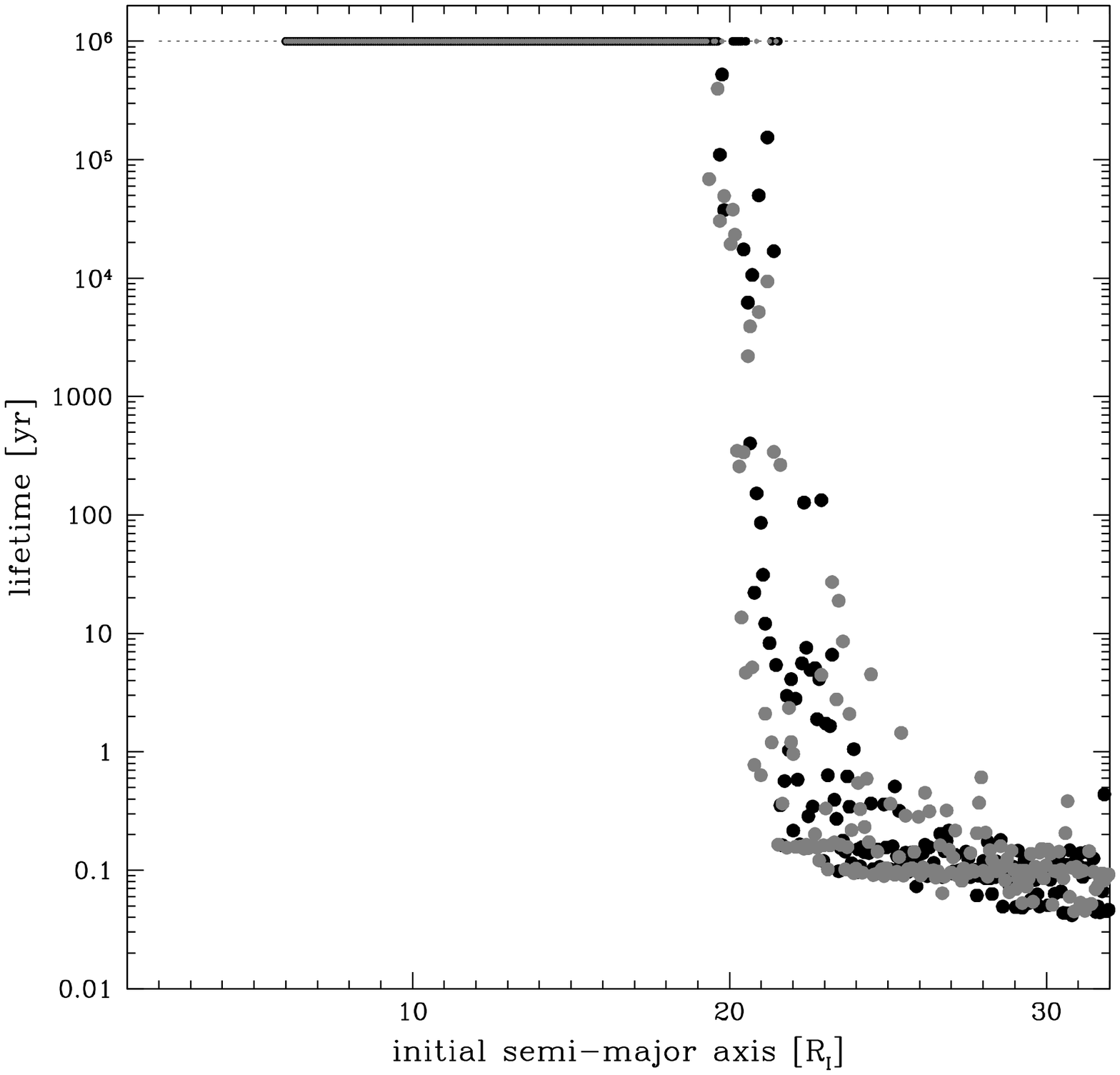}
\caption{The lifetime of each test particle is plotted as a function
  of their initial semimajor axis for two different intial
  eccentricities (black) 0.1 and (gray) 0.0. The simulations lasted
  for 1 Myr, which is shown as a horizontal line.  Symbols for
  particles which survive for 1 Myr are smaller than those of
  particles with shorter lifetimes. The lifetime drops precipitously
  at $a$ = 21~R$_{\mathrm{I}}$ = 0.44~R$_{\mathrm{H}}$.  }
\label{f1} %
\end{figure}

The simulations were performed in an Iapetus-centered frame.  The
lifetime of particles dropped precipitously beyond
0.4~$R_{\mathrm{H}}$, suggesting that any sub-satellite with a larger
semi-major axis would very quickly go into orbit around Saturn
(Fig.~\ref{f1}). Thus, we adopt this limit, which is equivalent to
21~$R_{\mathrm{I}}$, in our calculations below.

[Fig.~\ref{f1}. here]

\section{Tidal evolution of Iapetus}
\label{sec:tide}

The de-spinning of Iapetus by Saturn has long been considered
problematic, because for nominal $Q/|k_2|$
($\sim\!10^5$), Iapetus should not have de-spun over the age of the
solar system \citep{Peale1977}. Starting with the assumption of
constant $Q/|k_2|$, and using the standard de-spinning
timescale (Murray \& Dermott 1999, eq 4.163),

\begin{equation}\label{domps}
\dot{\Omega}_{\mathrm{I}} = -\mathrm{sign}(\Omega_\mathrm{I} - n)\frac{3|k_2|}{2\alpha Q}\frac{m_{\saturn}^2}{m_\mathrm{I}(m_{\saturn}+m_{\mathrm{I}})} \left(\frac{R_{\mathrm{I}}}{a}\right)^3n^2
\end{equation} 

\noindent where $\alpha \le$~2/5 is the moment of inertia constant of
Iapetus, $m_{\mathrm{I}}$ is its mass, $\Omega_{\mathrm{I}}$ is its
spin frequency, $|k_2|$ is the magnitude of the $k_2$ Love number, $Q$
the tidal dissipation factor, $m_{\saturn}$ is the mass of Saturn, and
$a$ and $n$ are the semi-major axis and mean motion of Iapetus.  The
$|k_2|$ and $Q$ values used throughout are for Iapetus only. For the
tidal interaction between Iapetus and Saturn, $\Omega_{\mathrm{I}} >
n$, so the effect is always to decrease the spin of Iapetus.

For these simple assumptions, the de-spinning from 16~h to a rate
synchronous with the orbital period, 79.3 days, takes
$3.6\times10^{5}\;(Q/|k_2|)$~years, nominally 36~Gyr, for a density
$\rho$ = 1~g~cm$^{-3}$. Using detailed geophysical models,
Castillo-Rogez et al.~(2007) and Robuchon et al.~(2010) showed Saturn
can de-spin Iapetus on solar system timescales, although only for a
narrow range of thermal histories.  Our goal here is to investigate
how the addition of the sub-satellite affects the de-spinning times.

Given that detailed models of Castillo-Rogez et al.~(2007) and
Robuchon et al.~(2010) used different methods, and that we are only
interested in how the de-spinning timescale changes with the addition
of a satellite, we take a simple approach of integrating a modified
version of Eq$.$~\ref{domps}. Our first adjustment is to remove the
assumption of constant $Q/|k_2|$.  This ratio is dependent on the
tidal frequency, ($\Omega - n$), and accounts for the manner in which
a material or body reacts to tidal stresses.  We start with a model of
Iapetus consisting of a time-invariant 200-km thick lithosphere with a
Maxwell viscoelastic rheology with rigidity $\mu=3.6\times10^9$ Pa and
viscosity $\eta$ = 10$^{22}$~Pa~s, which is strong enough to support
the equatorial bulge and ridge \cite{CR2007}, overlying a mixed
ice/rock mantle with a lower viscosity, representative of an interior
warmed by radiogenic heating.  We performed two types of simulations.
In the first, the viscosity of the mantle is held constant with time
and has values from $\eta$ = 10$^{15}$-- 10$^{18}$~Pa~s (typical for
the interior of an icy satellite at 240 -- 270 K).  In the second, we
allow $\eta$ of the inner ice/rock mantle to vary according to the
thermal evolution models in Castillo-Rogez et al.~(2007) and Robuchon
et al.~(2010).  In particular, we employ the LLRI model of
Castillo-Rogez et al.~(2007), and the 0.04 and 72~ppb $^{26}$Al cases
from Robuchon et al.~(2010).  Love numbers are calculated for a
spherically symmetric, uniform-density Iapetus using the SatStress
software package (Wahr et al.~2009).  We calculate the Love number
$k_2$ (which is a complex number for a viscoelastic body, see Wahr et
al., 2009 for discussion) and estimate $Q/|k_2|=1/{\rm Im}(k_2)$
(Segatz et al.~1988).  The values of $Q/|k_2|$ vary over an order of
magnitude for each value of $\eta$ for the important range of tidal
frequencies.

An integration of equation (\ref{domps}) was performed using a
Bulirsch-Stoer integrator for times up to 100 Gyr, incorporating the
frequency dependent $Q/|k_2|$ for different internal viscosities
which, in turn, is a function of temperature. Without the
sub-satellite, the time for Iapetus to reach synchronous rotation
ranged between 5~$\times10^8$ (fixed $\eta$~=~10$^{15}$~Pa~s) to
2~$\times10^{12}$ years (0.04~ppb $^{26}$Al case from Robuchon et
al.~2010).  We describe an investigation of the effect that a
sub-satellite could have on the spin of Iapetus in the next
subsection.

\subsection{Tidal interaction with a sub-satellite }
\label{ssec:tideSS}

The sub-satellite raises a tidal bulge on Iapetus, causing Iapetus to
de-spin and the sub-satellite's orbit to change.  The change in spin
rate for Iapetus due to a sub-satellite is (Murray \& Dermott 1999,
eq. 4.161),

\begin{equation}\label{dompss}
\dot{\Omega}_{\mathrm{I}} = -\mathrm{sign}(\Omega_\mathrm{I} - n)\frac{3|k_2|}{2\alpha Q}\frac{m_{\mathrm{ss}}^2}{m_{\mathrm{I}}(m_{\mathrm{I}}+m_{\mathrm{ss}})} \left(\frac{R_{\mathrm{I}}}{a}\right)^3n^2
\end{equation} 

\noindent and, the change in the satellite's orbit by (Murray \&
Dermott 1999, eq. 4.162),
\begin{equation}\label{dadt}
\dot{a} = \mathrm{sign}(\Omega_\mathrm{I} - n)\frac{3|k_2|}{2\alpha Q}\frac{m_{\mathrm{ss}}}{m_{\mathrm{I}}}\left(\frac{R_{\mathrm{I}}}{a}\right)^5na\; .
\end{equation} 

\noindent Together Eqs.~\ref{dompss} and \ref{dadt} describe the
interaction between the sub-satellite and Iapetus, where
$m_{\mathrm{ss}}$ is the mass of the sub-satellite. The term
$\mathrm{sign}(\Omega_\mathrm{I} - n)$ is of great importance,
determining whether the satellite evolves outward while decreasing the
spin of Iapetus, or inwards while increasing the spin of Iapetus.  At
semi-major axis $a_{\mathrm{sync}} =
(G(m_{\mathrm{I}}+m_{\mathrm{ss}})/\Omega_\mathrm{I}^2)^{3/2}$,
$\Omega_\mathrm{I} = n$, representing a synchronous state. If the
sub-satellite has $a<a_{\mathrm{sync}}$, it evolves inwards; if
$a>a_{\mathrm{sync}}$, it evolves outwards.  Saturn is gradually
decreasing the rotation rate of Iapetus, and thus the synchronous
limit slowly grows larger, possibly catching and overtaking a slowly
evolving sub-satellite. The orbital period of a sub-satellite at
21~$R_{\mathrm{I}}$, the distance at which we consider a satellite
stripped by Saturn, is $\sim$~12.8~days.  Thus, if Iapetus is de-spun
to a period of 12.8~days before the sub-satellite reaches
21~$R_{\mathrm{I}}$, it will be caught by the expanding synchronous
limit.

For the integrations of the sub-satellite's tidal evolution, the
sub-satellite's mass is used as a free parameter, while the starting
semi-major axis is set to 3~$R_{\mathrm{I}}$. This distance is derived
from the expected origin of the sub-satellite accreting from an
impact-caused debris disk encircling Iapetus \citep{Ida97,Kokubo00}.
However, tidal evolution timescales for a sub-satellite are largely
insensitive to the initial semi-major axis, so this starting point
only needs to be beyond the synchronous limit for the model to be
accurate.  With a rotation period of 16~h for Iapetus,
$a_{\mathrm{sync}}$ would have been 2.94~$R_{\mathrm{I}}$, which is
outside the Roche limit defined to be at $r_{\mathrm{roche}} = 2.46
R_{\mathrm{I}}(\rho_{\mathrm{I}}/\rho_{\mathrm{ss}})^{(1/3)} \approx
2.53 R_{\mathrm{I}}$ for $\rho_{\mathrm{ss}} = 1$ g cm$^{-3}$. Thus, a
satellite forming at 3.0~$R_{\mathrm{I}}$ would be above the
synchronous limit and destined, initially, to evolve outward due to
tidal interaction with Iapetus.

Equations (2) and (3) were then integrated with equation (1), to
follow the evolution of the spin of Iapetus due to Saturn and the
sub-satellite. We studied the geophysical models for Iapetus described
in the last section, along with sub-satellites with mass ratios, $q
\equiv m_{\mathrm{ss}}/m_{\mathrm{I}}$, between 0.0001 and 0.04.  We
summarize the suite of simulations in Fig. 2, showing the time of
de-spinning for Iapetus and the time at which the sub-satellite is
stripped by Saturn or tidally evolves back to re-impact Iapetus.  We
present the data scaled to the de-spinning time due to Saturn alone to
highlight the effect of the sub-satellite in accelerating the tidal
evolution of the system.  The fate of the system, in regards to the
escape or re-impact of the sub-satellite and the de-spinning time of
Iapetus, are separated into three distinct classes of outcomes based
on $q$.

\subsubsection{Synchronous lock and re-impact: $q > 0.021$}

Above a mass ratio $q > 0.021$, the sub-satellite does not reach
$a_{\mathrm{st}}$ before becoming synchronous with the spin of Iapetus
(see Fig.~\ref{frames}a). With both the sub-satellite and Saturn
working to slow the spin of Iapetus, the synchronous limit grows to
21~$R_{\mathrm{I}}$ before the sub-satellite evolves to that
semi-major axis. This result only varies mildly for the different
geophysical models of Iapetus, as all three timescales depend linearly
on $Q/|k_2|$; therefore, the re-impact outcome only depends on mass
ratio. However, the time to reach this outcome ranges from $10^{12}$
yr for the Robuchon et al.~(2010) models to $10^9$ yr for the
Castillo-Rogez et al.~(2007) LLRI model.

Upon reaching synchronous rotation, the evolution does not stop
because Saturn is still tidally interacting with Iapetus.  As Saturn
continues to slow the spin rate of Iapetus, the synchronous limit
moves beyond the sub-satellite, which then begins to tidally evolve
inwards. The sub-satellite is doomed to evolve inwards and hit
Iapetus.  Given that the sub-satellite started at 3~$R_{\mathrm{I}}$
and finishes by impacting Iapetus, it makes a net contribution to the
angular momentum of Iapetus, and so it is spinning faster than if the
sub-satellite were never there. Thus, the de-spinning of
Iapetus (after re-impact) finishes later than it would have by Saturn
tides alone (see Fig.~\ref{frames}a).

\subsubsection{Satellite is stripped: $0.006 < q < 0.021 $}\label{313}

For $0.006 < q < 0.021$, the sub-satellite evolves to
21~$R_{\mathrm{I}}$ and is stripped by Saturn before attaining a
synchronous orbit.  As it moves out, the sub-satellite carries away
angular momentum from Iapetus, allowing it to rapidly de-spin.  This
angular momentum is then removed from the Iapetus system when the
sub-satellite is stripped.

The sample evolution for a system with a constant $\eta = 10^{16}$ Pa
s and $q=0.018$ (Fig.~\ref{frames}b) shows that the sub-satellite
reaches an orbital period of $\sim$~12 days ($a =21\;R_{\mathrm{I}}$)
before Iapetus reaches that spin period.  It is important to note that
the sub-satellite does not totally de-spin Iapetus.  Instead, it slows
Iapetus down enough that Saturnian tides (which are faster because
$Q/|k_2|$ is a decreasing function of the spin rate) can finish the
job.  For all but one (see \S{\ref{ssec:discuss}}) of our geophysical
models, the sub-satellite can de-spin Iapetus an order of magnitude
faster than it is when de-spun by Saturn alone.  This order of
magnitude difference means that Iapetus could have de-spun in 500 Myr,
for situations that would otherwise require the age of the solar
system.

\subsubsection{Slow evolution of a small satellite: $q<0.006$}

The tidal evolution timescale of the sub-satellite's orbit expansion
depends on $q$, and for smaller mass ratios, the evolution takes
longer. Below $q<0.006$, the de-spinning of Iapetus due to Saturn is
fast enough that the location of synchronous rotation sweeps past the
sub-satellite (see Fig.~\ref{frames}c). After this occurs, the
sub-satellite is then below the synchronous limit and doomed to evolve
back in towards Iapetus. In this scenario, the evolution of the
sub-satellite back to the surface of Iapetus takes longer than it
takes for Iapetus to de-spin.

Iapetus de-spins faster than Saturn otherwise could do alone.  In this
case, however, the relevant time constraint becomes the sub-satellite
impact time (the small open symbols in Fig.~\ref{alltides}) rather
than the de-spinning time because we currently do not see a satellite.
We find that sub-satellite impact time can be shorter than the
de-spinning time for $q>0.003$.  However, it should be noted that this
dynamical pathway only helps by at most roughly a factor of 2 over
Saturn acting alone.  In addition, the sub-satellite is likely to
tidally disrupt on its way in, forming a second, significantly
fresher, ridge.  This is probably inconsistent with the ridge's
ancient appearance.  Thus, we think that this particular dynamical
pathway can probably be ruled out, but we include it for completeness.

\subsubsection{Discussion of sub-satellite tides} 
\label{ssec:discuss}

The integrations have bracketed the possible behavior of the
Saturn-Iapetus-sub-satellite system. At high and low mass ratios the
sub-satellite is doomed to return to re-impact Iapetus, while for
$0.006 < q < 0.021$ the sub-satellite is stripped.  As
Fig.~\ref{alltides} shows, sub-satellites with masses between $0.005 <
q < 0.021$ decrease the despinning time over that of Saturn alone.
This effect can be as large as a factor of 10 for $q\sim 0.02$.
Consistent with prior work, we find that a low-viscosity interior
(presumably warmed by radiogenic heating) is required to despin
Iapetus over solar system history.  We find that the age of the
despinning event with and without a sub-satellite are similar if the
thermal evolution of Iapetus follows the LLRI model of Castillo-Rogez
et al.~(2007).  In this case, Iapetus' interior heats slowly during
the first Gyr of its history.  When the interior is warmed close to
the melting point, tides drive rapid despinning. The presence of the
sub-satellite can significantly shorten this period of time, but
because the de-spinning time is short irrespective of whether the
sub-satellite is present, it does not significantly alter the time in
solar system history when Iapetus is de-spun.


\section{The ridge and sub-satellite formation}

Given the sub-satellite masses which can assist in de-spinning Iapetus,
and estimates on the mass of the equatorial ridge, a constraint can be
placed on the amount of mass placed into orbit by the original
disk-forming impact. The ring of debris which collapses to form the
ridge must do so rapidly, as a ring is not currently observed, and the
ridge is one of the oldest features on the surface of Iapetus (Giese
et al. 2008).

\subsection{Ridge mass}
\label{sec:RM}

The ridge has an unknown mass due to incomplete imaging and
significant damage from cratering. Ip (2006) estimated its mass
assuming that it had, at one time, completely encircled the equator
with a height of 20 km and width of 50 km, $m_{\mathrm{ridge}} = 2
\times \pi \times R_{\mathrm{I}}\times 20\: \mathrm{km} \times 50\:
\mathrm{km}$ with a density of 1 g cm$^{-3}$ equals a mass of
4.4$\times10^{21}$ grams (Ip 2006 used a radius of 713 km for this
mass estimate).  Giese et al.  (2008) found a maximum height of 13 km
in Digital Terrain Models (DTM) models, though the shape model maximum
height was 20 km.  Comparing the dimensions given by Ip (2006) with
the profiles in Giese et al.  (2008), we set a lower limit by taking a
factor of two in both vertical and horizontal extent and assuming that
the cross section is a triangle rather than a rectangle, yielding an
estimate $\sim$8 times lower, 5.5$\times 10^{20}$ grams (where we use
a radius of 746 km). However, Castillo-Rogez et al. (2007) quote ridge
dimensions of 18 km by 200 km over a length of 1600 km, which yields a
mass of 3$\times10^{21}$ grams, similar to the Ip (2006) estimate.
Profiles from the DTM model (Giese et al. 2008; Fig.~5) do not show a
recognizable base greater than 50 km for 10 different profiles across
the ridge and no elevations greater than $\sim$13~km, making this a
very high estimate.

We assume that the ridge consists of ring material that lands on the
surface of Iapetus, accounting for the equatorial location.  Given a
ring of material interior to the Roche limit, there are two ways for
it to land on the surface of Iapetus: the ring can tidally evolve down
to the surface, or it can be pushed there by the newly formed
sub-satellite. For reasons discussed below, we focus on the latter.

\subsection{Sub-satellite and ring interaction}
\label{ssec:ring}

The tidal evolution of the ring down to the surface of Iapetus
requires that the material be inside the synchronous rotation height.
As described above, the synchronous height for a rotation period of
16~h is at $a=2.94 $~R$_{\mathrm{I}}$, which is exterior to the Roche
limit at $r_{\mathrm{roche}} = 2.53 R_{\mathrm{I}}$ for
$\rho_{\mathrm{ss}} = 1$ g cm$^{-3}$. The ring material evolves due to
tides with Iapetus and the sub-satellite that accretes beyond the
Roche limit.  By comparing our estimates for the sub-satellite
(\S{\ref{ssec:tideSS}}) to those of the ridge ({\S{\ref{sec:RM}}}), we
find that the sub-satellite is more massive than the ridge for the
entire range of sub-satellite masses that assist in de-spinning.  In
this case, the ring spreading timescale is the time it takes a
particle to random walk across a distance $r$ \cite{Goldreich1982},

\begin{equation}
  t_{\mathrm{spread}} =(191 \:\mathrm{yr})\left(\frac{5600\: \mathrm{g \: cm^{-2}}}{\Sigma}\right)\left(\frac{150\: \mathrm{km}}{R_{\mathrm{ss}}}\right)^3\left|\frac{a_{\mathrm{ss}} - r}{566\: \mathrm{km}}\right|^3,
\end{equation}
where $\Sigma$ is the surface density of the ring, and
$R_{\mathrm{ss}}$ and $a_{\mathrm{ss}}$ are the radius and the
semi-major axis of the sub-satellite, respectively. The surface
density ($\Sigma$) of the ring is simply the ridge mass spread over
the region interior to the Roche limit, $\sim5.6$--$46.4\times10^3$ g
cm$^{-2}$. The possible range of sub-satellite masses is equivalent to
the mass of a single body of radius, $R_{\mathrm{ss}}$, of
$\sim$131--211~km for a density of 1~g~cm$^{-3}$ (or 155--251~km for
$\rho_{\mathrm{ss}}$=0.6~g~cm$^{-3}$). The semi-major axis of the
sub-satellite is likely to be 1.3~$R_{\mathrm{roche}}$ initially
\citep{Kokubo00}, and so the range of possible times for the spreading
of the ring is 9--286~years. Thus, even with many conservative
approximations, this timescale is many orders of magnitude shorter
than other timescales in the problem.

For the sub-satellite masses of interest, the effect of the ring on
the sub-satellite would be dwarfed by the much larger effect of
Iapetus's tides.

\section{Impact scenarios}
\label{sec:impact}

In the vast majority of situations in which the addition of a
sub-satellite aids de-spinning, the sub-satellite is stripped from its
orbit around Iapetus. In these cases the stripped sub-satellite will
still be bound in the Saturnian system, at least initially.  In this
section, we determine the possible fates for these objects and ask
what effect this will have on Iapetus.

\subsection{The dynamical fate of a stripped sub-satellite}
\label{ssec:fate}

To determine the probability of impact by a stripped sub-satellite we
performed a numerical $N$-body experiment consisting of the orbital
evolution of 50 massless test particles initially in orbit around
Iapetus. We used SyMBA, a symplectic code which is capable of handling
close encounters \cite{Duncan98}. The simulations included Titan,
Hyperion, Iapetus, and Phoebe.  Particles were stopped if they hit a
satellite, crossed Titan's orbit, became unbound from Saturn, or
reached a distance of 0.4 AU from Saturn (roughly its Hill radius).
The particles' initial semi-major axes ranged from 0.4 to 0.8
$R_{\mathrm{H}}$ (19--$38 R_{\mathrm{I}}$) with eccentricities of 0.1.
The particles' inclinations were between 0$^{\circ}$ and 15$^{\circ}$
degrees with respect to Saturn's equator.

After 6 Myr only 1 particle remained in orbit around Iapetus, while 3
particles hit Iapetus before becoming unbound. The remaining 46
particles became unbound from Iapetus, entering orbit about Saturn.
These are the objects of interest here because our goal is to
determine the fate of a sub-satellite {\it once} it becomes unbound.
Five of them were ejected from the Saturn system by the satellites
(entering heliocentric orbit).  The remaining 41 objects impacted
Iapetus --- none hit Titan, Hyperion, or Phoebe.  Thus, a stripped
sub-satellite has a roughly $41/46 \sim 90$\% chance of ending its
existence by returning from orbit about Saturn and colliding with
Iapetus.  This fact means that we must consider the effects of such an
impact in our scenario.

\subsection{Angular Momentum Budget}

In cases in which the stripped sub-satellite re-impacts Iapetus, we
must consider the angular momentum it imparts. If the resulting spin
rate is too fast it will cancel any advantage that was originally
gained by the presence of the sub-satellite.  We assume that our
scenario is still viable if, after the impact, $\Omega_\mathrm{I} <
\Omega^\ast = 2 \times 10^{-5}$, where $\Omega_\mathrm{I}$ is the spin
frequency of Iapetus.  For spin rates $< \Omega^\ast$, corresponding
to spin periods greater than $\sim\!4$ days, the de-spinning time after
the impact will be less than $\sim 100$~Myr for $\eta$=10$^{16}$~Pa~s.

Assuming the sub-satellite is accreted completely, the magnitude of
the angular momentum brought in by the sub-satellite is $H =
m{_\mathrm{ss}} v_\infty b$, where $v_\infty$ is the satellite's
velocity with respect to Iapetus at ``infinity'', and $b$ is the
impact parameter. Assuming that Iapetus's pre-impact rotation is slow,
we find that $\Omega_\mathrm{I} < \Omega^\ast$ requires that $b <
b^\ast \equiv 2 R_\mathrm{I}^2 \Omega^\ast/ 5 q v_\infty$. The maximum
value of $b$ that allows a collision with Iapetus is $b_{\rm max}^2 =
R_\mathrm{I}^2 [1 + (v_{\rm esc}/v_\infty)^2]$, where $v_{\rm esc}$ is
Iapetus' surface escape speed. If $b^\ast > b_{\rm max}$, all impacts
leave Iapetus spinning more slowly than $\Omega^\ast$; otherwise the
probability $P$ that Iapetus will have $\Omega_\mathrm{I} <
\Omega^\ast$ (and thus our scenario will remain viable) is
$(b^\ast/b_{\rm max})^2$ = $(2 R_\mathrm{I} \Omega^\ast/5qv)^2$,
where $v$ is the impact speed and we have used $v^2 = v_{\rm esc}^2 +
v_\infty^2$.

In \S{\ref{ssec:tideSS}}, we found that satellites with $0.005 < q <
0.021$ were most effective in de-spinning Iapetus. Most impacts
occurred at velocities near the escape speed of Iapetus, 0.58 km/s.
At that speed, $P = 1$ for $q < 0.0103$ and $P = (0.0103/q)^2$ for
$0.0103 < q < 0.021$. Thus, the lower-mass sub-satellites never
produce spin rates faster than $\Omega^\ast$, while sub-satellites
with $q = 0.0146$ yield viable scenarios 50\% of the time. We
therefore conclude that the re-impact of the sub-satellite can be
consistent with Iapetus's spin state.

\subsection{Linking basins to the possible sub-satellite impact}

A large complex crater or basin will form if the sub-satellite were to
impact Iapetus. In this section, we estimate the size of the impactors
needed to produce the basins observed on Iapetus today and compare
them to the size of the sub-satellite.  Zahnle et al$.$ (2003)
estimate the diameter, $D$, of the final (collapsed) crater on a
mid-sized icy satellite to be
\begin{equation}
D = 21.4 \, {\rm km} \left({v \over 1\,{\rm km/s}}\right)^{0.49}
\left({1\,{\rm cm/s^2} \over g}\right)^{0.245}
\left({\rho_i \over \rho}\right)^{0.377} 
\left({R_{ss} \over 1\,{\rm km}}\right)^{0.885},
\end{equation}
where $v$ is the normal component of the impact velocity, $g$ is the
surface gravity of the target, $\rho$ is the density of the target,
and $\rho_i$ is the density of the impactor. This equation assumes
that the incidence angle of the impact, measured from the normal to
the target, is $45^\circ$.  If we substitute $g = 22.3$~cm s$^{-2}$
for Iapetus, $\rho_i = \rho = 1.0$~g~cm$^{-3}$, we have
\begin{equation}
\label{eq:scale}
D = 589 \, {\rm km} ~~~ \left({v \over 1\,{\rm km/s}}\right)^{0.49}
\left({R_{ss} \over 100\,{\rm km}}\right)^{0.885}.
\end{equation}

Giese et al. (2008) found 7 basins, defined as craters with $D >
300$~km, on the leading face of Iapetus. The largest is stated to have
$D =$~800 km. The Gazetteer of Planetary Nomenclature ({\tt
  http://planetarynames.wr.usgs.gov/}) lists 5 basins on Iapetus, with
the largest, Turgis, 580 km in diameter. 

In our scenario, one of Iapetus's basins may have been created by our
escaped sub-satellite when it returns to meet its maker.  In our
simulations in \S{\ref{ssec:fate}}, we find that the impact would
likely occur at a velocity near Iapetus's escape velocity, 0.58 km/s;
however, there is about a 10\% chance that the impact speed would
exceed 2~km/s. At 2~km/s, Eq.~(\ref{eq:scale}) indicates that 300 --
800 km basins are produced by impactors with radii between 32 and
96~km. At the more likely speed of 0.58~km/s, the corresponding
impactor radii are 64 and 193~km.  Recall that in \S{\ref{ssec:ring}},
we found that our hypothetical sub-satellite should have radii between
131 and 211~km for these densities. The ranges vary slightly for a
sub-satellite of only 0.6 g cm$^{-3}$ with impactors between 40--120
km at 2 km/s and 80--241 km at 0.58 km/s, where the hypothetical
sub-satellite at this density would have a radii between 155--251 km.
Therefore, it is quite possible that one of the observed basins was
caused by our hypothetical sub-satellite.

These calculations are highly uncertain because numerical simulations
at the relevant scales and velocities have not, to our knowledge, been
performed for icy targets. In particular, scaling relations in the
literature, such as the Schmidt-Housen scaling we used above, are
generally based on field data or simulations of hypervelocity impacts,
i.e., impacts at velocities large compared with the speed of sound in
the target, which is about 3 km/s for non-porous ice. Impacts by
Iapetus's putative sub-satellite would have occurred at lower speeds.
Furthermore, experiments and simulations generally deal with impactors
that are much smaller than their targets. This condition is only
marginally satisfied at typical speeds for the impacting sub-satellite
in our scenario. Thus, the values we quote for the sub-satellite's
size should be viewed as rough estimates.  However, the calculations
do support the possibility of basin formation caused by the impacting
sub-satellite.


\section{Discussion and conclusions}

We have explored the scenario in which a sub-satellite forms from an
impact-generated debris disk around Iapetus following an impact. The
remains of the disk fall to the surface of Iapetus to build the
observed equatorial ridge, while the tidal evolution of the
sub-satellite assists in de-spinning Iapetus.  We find that this
scenario can significantly shorten Iapetus's tidal de-spinning time if
the mass ratio between the sub-satellite and Iapetus, $q$, is between
0.005 and $0.021$. These results suggest that the presence of a
sub-satellite can potentially loosen constraints on the geophysical
history of Iapetus that have been implied by the timing and duration
of despinning.  The full implications of our results, however, cannot
be realized until the sub-satellite scenario is investigated using a
detailed thermal evolution model such as those described in prior
works (e.g$.$ Castillo-Rogez et al.~(2007); Robuchon et al.~(2010)).

Iapetus has been of great interest due to its extremely old surface
that records the cratering history of the outer solar system.  Prior
scenarios require that the ridge form after an epoch of early heating
to de-spin Iapetus that drive the timing and duration of resurfacing
(Castillo-Rogez et al.~(2007); Robuchon et al.~(2010)). In our model,
the ridge forms only a few hundred years after the impact, and
therefore the ridge is quite old. This matches the qualitative
assessment that the ridge is one of the oldest features on the
surface, along with the 800-km basin \citep{Giese2008}.  Some of the
sub-satellite evolutionary scenarios presented here end with a
re-impact, requiring an associated basin forming much later.

The ridge is only seen to extend roughly $110^\circ$ around Iapetus.
This could be the result of its extreme age --- much of it could have
been destroyed by subsequent cratering.  Thus, it is still unclear
whether the ridge extends for the full extent of the equatorial
circumference, but an infalling ring would preferentially deposit
material on global topographic high terrain.  Indeed, a ring that
extends only $110^{\circ}$ in longitude could result if the
center-of-figure of Iapetus were offset from its center-of-mass as is
seen on the Moon \citep[see][and references therein]{Araki09}.

\begin{figure}
\includegraphics*[width=6in]{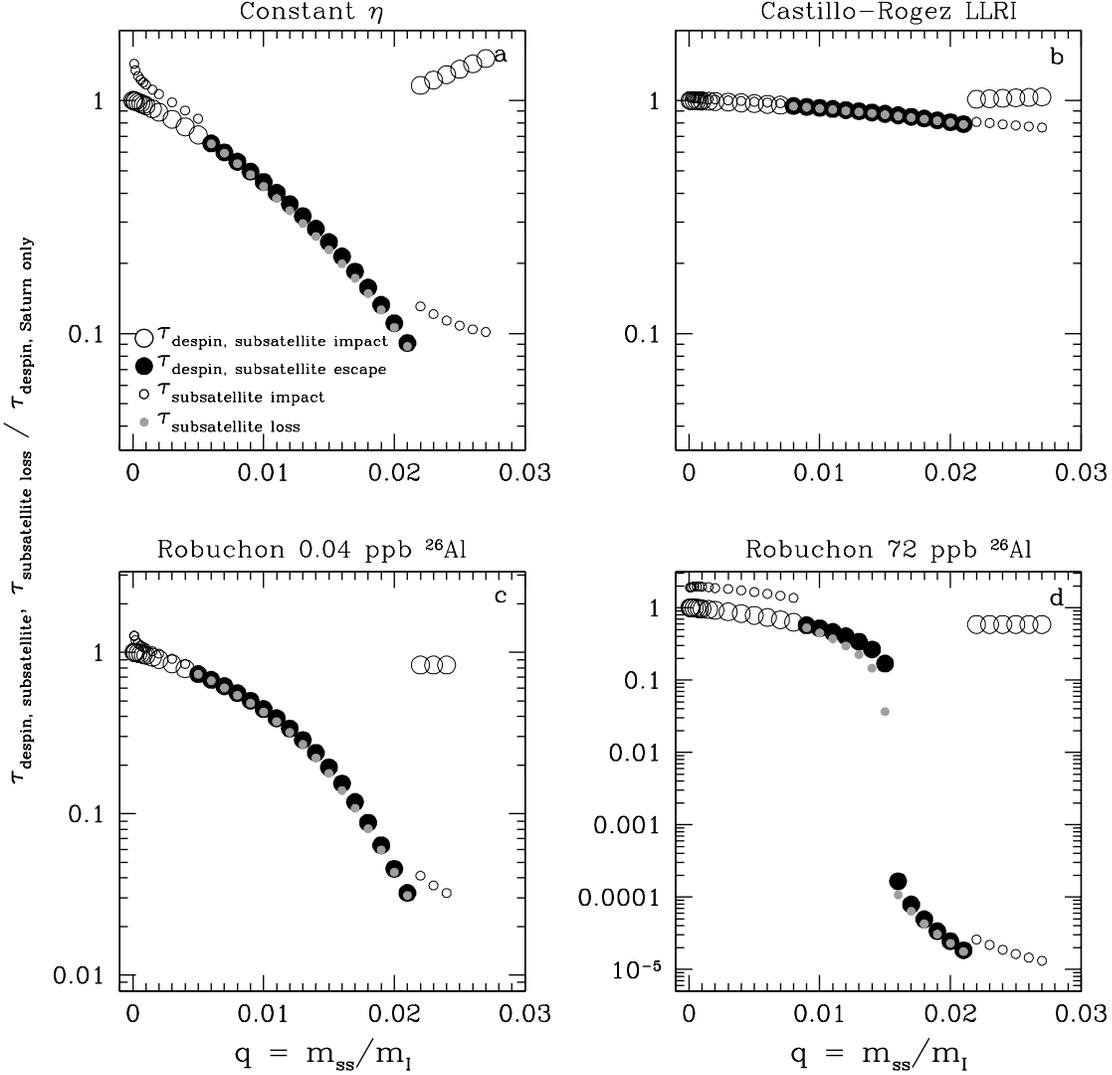}
 \caption{Results from the integrations of the despinning of
   Iapetus, where despinning and sub-satellite stripping/impacting
   timescales were calculated for different mass ratios $q$ for
   different internal viscosities $\eta$ and different internal
   temperature evolution. The filled symbols are the despinning times
   for simulations where the sub-satellite is stripped, where the
   smaller symbol at the same $q$ is the time of stripping. The open
   symbols are for cases where the sub-satellite evolves back towards
   Iapetus, and the smaller symbol marks the time of re-impact. The
   times are normalized to the time it takes Iapetus to de-spin by
   Saturnian tides alone for each case. Four cases are plotted: (a)
   constant viscosity, where the results from each constant value of
   $\eta=10^{15}$ Pa s, $\eta=10^{16}$ Pa s, etc.  plot on top of each
   other when normalized to 1.0, (b) the LLRI case from Castillo-Rogez
   (2007), (c) the 0.04 ppb and (d) 72 ppb $^{26}$Al cases from
   Robuchon et al. (2010).}
 \label{alltides} %
\end{figure}

\begin{figure}
\includegraphics*[width=6in]{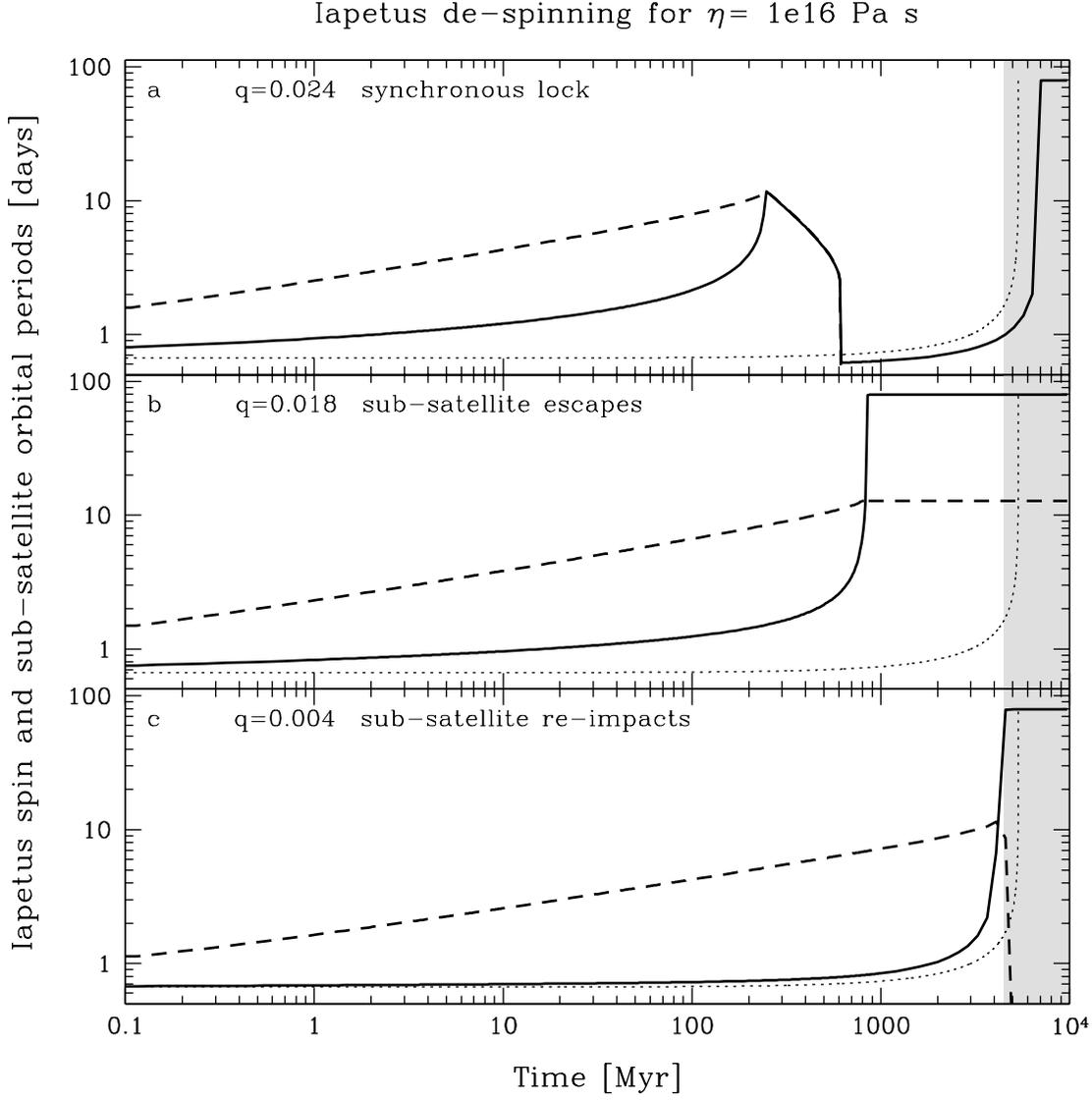}
\caption{Evolution of the spin period of Iapetus (solid line) and
  orbital period of the sub-satellite (dashed line) as a function of
  time, for with constant $\eta$ equal to $10^{16}$ Pa s and mass
  ratios of (a) $q=0.024$, (b) $q=0.018$ and (c) $q=0.004$.  The
  dotted line in each panel is the evolution of the spin of Iapetus
  under the influence of Saturn only, and the shaded region marks the
  age of the solar system. They illustrate three different outcomes.
  In case (a) the sub-satellite is too large and eventually is caught
  in synchronous lock with the rotation of Iapetus. Saturn continues
  despinning Iapetus, and so the sub-satellite falls below synchronous
  height, returning to impact Iapetus. Iapetus then despins, again,
  due to Saturn, and finally reaches a despun state later than had it
  simply despun due to the effects of Saturn. In case (b) the
  sub-satellite assists in despinning Iapetus, and is then stripped by
  Saturn, allowing Iapetus to despin up to 10$\times$ faster than by
  Saturn alone.  Finally, in (c), the small sub-satellite's orbit
  evolves very slowly, so that Iapetus is despun by Saturn fast enough
  for the synchronous limit to move beyond the sub-satellite, forcing
  the sub-satellite to tidally contract its orbit and return to impact
  Iapetus. For this case the despinning time is similar to that by
  Saturn's effect alone. }
\label{frames} %
\end{figure}

The stripped sub-satellite mass range, described above, corresponds to
bodies with radii of between roughly 130 and 210~km (assuming
$\rho_{\mathrm{I}}$ =1~g~cm$^{-3}$), not all of which are below the
estimated radius, $\sim$190~km, of the largest allowable impactor
given the basins on Iapetus.  This limiting size corresponds to a mass
ratio $q$ = 0.015.  We arrive at a similar upper limit when
considering the angular momentum of the impact.  Thus, to account for
the impact of the secondary, the mass ratio range 0.006$ < q <$0.015
is required. As seen in Fig.~\ref{alltides}, the sub-satellites are
stripped on very similar timescales to the de-spinning of Iapetus.
Thus, a sub-satellite returning to form a basin on Iapetus would do so
after the ridge formed.  This younger basin caused by the
sub-satellite would then be expected to overlie the ridge if it was a
near-equatorial impact.  Turgis (the 580~km Basin II in Giese et al.
2008) fulfills these requirements as the equatorial ridge appears to
stop when intersecting this basin (though there are no profiles
presented in Giese et al$.$~2008 for this region of the surface). This
basin would correspond to an impactor with a radius of 140~km, which
is only slightly larger than the lower limit estimated above.
However, we find this acceptable given the inherent uncertainties in
calculating crater sizes.  Meanwhile Basin I, the largest at 800~km,
appears to have a stratigraphically similar age as the ridge
\citep{Giese2008}.  It is important to note, however, that our
simulations predict only $\sim$90\% of stripped sub-satellites return
to impact, so it is not a certainty that one of the basins is
associated with this evolution.

A strength of our hypothesis is that the limits on impactor and
sub-satellite masses ($0.005 < q < 0.015$) are in line with the
estimated ridge mass and the number of 300--800~km basins. Thus, this
work provides a complete story from original to final impact which may
explain the ridge, shape, and basin population on Iapetus.

\acknowledgments We would like to thank Michelle Kirchoff and David
Nesvorn{\' y} for useful discussions.  HFL and KJW are grateful for
funding from NASA's Origins and OPR program.  ACB and LD acknowledge
support from NASA CDAP grants.




\newpage

\newpage

\end{document}